# Linewidth of collimated wavelength-converted emission in Rb vapours

Alexander Akulshin[1], Christopher Perrella[2], Gar-Wing Truong[2], Andre Luiten[2,3], Dmitry Budker[4], and Russell McLean[1]

[1] Centre for Atom Optics and Ultrafast Spectroscopy, Swinburne University of Technology, Melbourne, Australia
[2] School of Physics, University of Western Australia, Nedlands 6009 WA, Australia
[3] Institute of Photonics and Advanced Sensing and the School of Chemistry and Physics, The University of Adelaide, Adelaide 5005 SA, Australia
[4] Department of Physics, University of California, Berkeley, CA 94720-7300, USA

E-mail: *aakoulchine@swin.edu.au*

**Abstract.** We present a study of the spectral linewidth of collimated blue light (CBL) that results from wave mixing of low-power cw laser radiation at 780 nm and 776 nm and an internally-generated mid-IR field at 5.23 μm in Rb vapour. Using a high-finesse Fabry-Perot interferometer the spectral width of the CBL is found to be less than 1.3 MHz for a wide range of experimental conditions. We demonstrate that the CBL linewidth is mainly limited by the temporal coherence of the applied laser fields rather than the atom-light interaction itself. Results obtained with frequency modulated laser light allow an upper limit of several hundred kHz to be set for the linewidth of the collimated mid-IR radiation at 5.23 μm, which has not been directly detected.

## 1. Introduction

Nonlinear parametric processes in atomic media can generate new optical fields with substantially different wavelengths [1]. A long-established and widely used parametric process is four-wave mixing (FWM), which was discovered using high-power pulsed lasers. However, it is also possible to drive this process with low-power, continuous-wave (cw) narrow-linewidth laser sources, allowing one to generate new highly monochromatic radiation.

Frequency conversion of low-power near-infrared laser light into blue and mid-infrared radiation using a FWM technique in Rb vapours was pioneered by Zibrov et al [2]. Since then it has become an active area of research of several groups [3,4,5,6,7] and has been even suggested for advanced undergraduate laboratories [8]. The technique is of interest because new optical field generation in atomic media can lead to important applications in quantum-information science [9] and low atom number detection [10]. Both of these possible applications exploit the potential to convert light from one spectral region to another. Furthermore, the same scheme, which is closely related to mirrorless lasing [11], can generate coherent and tuneable radiation at wavelengths near-resonant to atomic transitions for which commercial light sources are not readily available.

The spectral properties of frequency up-converted radiation have been briefly considered in [2, 3]. A study of the tunability and absolute frequency of the CBL generated using a diamond-shaped transition configuration (Fig.1a) was presented in [12] along with schemes for its frequency and intensity stabilization, but a systematic investigation of the linewidth of the frequency-converted radiation has not been reported to date. In this work, we use a high-resolution Fabry-Perot interferometer (FPI) to make such a study on the CBL at 420 nm generated by parametric wave mixing in warm Rb vapours under similar conditions to those of [4, 12]. The results obtained allow an upper limit to be set for the linewidth of mid-IR radiation that was not directly detected in our experiment. Although we have used Rb vapours, we expect the results to be applicable to similar wave-mixing schemes and other alkali atoms.

## 2. Experimental set-up

A simplified diagram of the Rb atomic energy levels used for CBL generation and the basic optical scheme of the experiment are shown in Fig. 1.

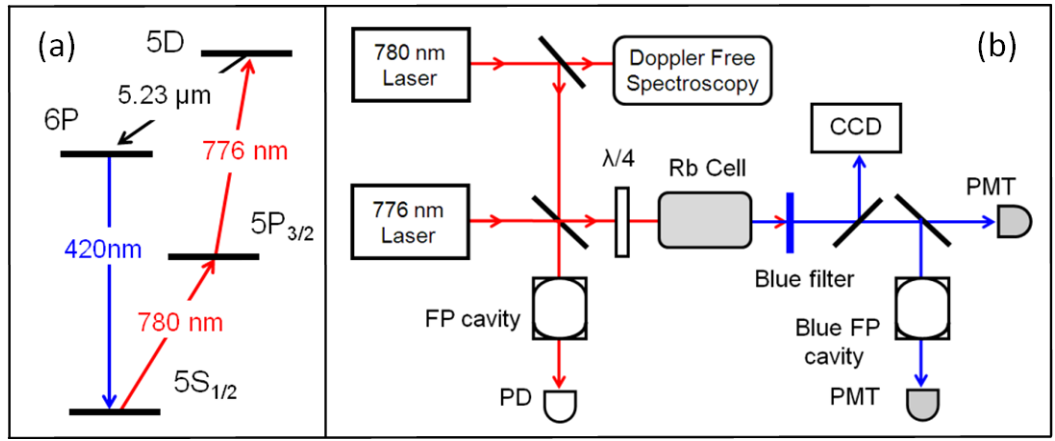

Figure 1. a) Diamond-shaped transition configuration for the case of Rb atoms. b) Simplified scheme of the experimental setup.

Extended-cavity diode laser (ECDL) sources at 780 and 776 nm are used for step-wise and two-photon excitation of Rb atoms to the $5D_{5/2}$ level. The optical frequency of the 780 nm ECDL is either scanned across the Rb $D_2$ absorption line or modulation-free stabilized to the Doppler-free polarization-spectroscopy resonance of the $^{85}$Rb $5S_{1/2}(F=3) \rightarrow 5P_{3/2}(F'=4)$ transition in an auxiliary Rb cell. The typical RMS optical-frequency fluctuations estimated from the error signal of the laser servo system are about 200 kHz over a one second time interval. The 776 nm laser is also either scanned across the $^{85}$Rb $5P_{3/2} \rightarrow 5D_{5/2}$ transition or modulation-free locked to a low-finesse tuneable reference cavity. When locked, frequency fluctuations of this laser are reduced to the 320 kHz level over the same timescale.

Radiation from both ECDLs is combined to form a bichromatic beam, the polarizations of the 780 nm and 776 nm components being controlled with wave plates and polarizers. The power of the 780 and 776 nm components is approximately 12 and 6 mW, respectively. The cross section of the beam inside the Rb cell is reduced by a long focal-length lens ($f$ =1 m) to about 0.5 mm$^2$. The temperature of the 5 cm long Rb cell, containing a natural mixture of Rb isotopes and no buffer gas, is set within the range 50-100°C, so that the atomic density $N$ of saturated Rb vapour varies between $1.2 \times 10^{11}$ cm$^{-3}$ and $6 \times 10^{12}$ cm$^{-3}$. A μ-metal shield is used to reduce the ambient magnetic field in the Rb cell to a few milligauss.

At various times in the experiment additional ECDLs tuned to different Rb transitions were used. A low-power, narrow-linewidth reference laser (FWHM < 200 kHz), was tuned to either the 776 or 780 nm spectral region to estimate the temporal coherence of the two drive lasers using a standard heterodyne method [13]. Another ECDL tuned to the Rb $D_1$ line at

795 nm was employed for ground-state hyperfine optical pumping to modify the efficiency of CBL generation, as demonstrated in [14].

Color filters of optical density approximately 0.5 and 4.0 at 420 nm and 780 nm respectively, were used to spectrally separate the CBL from the laser radiation. We detect the filtered blue light using photomultipliers (PMTs) and CCD cameras. The CBL spectral purity and linewidth were explored using an optical analyser (ANDO AQ-6315E) and a tuneable Fabry-Perot interferometer (FPI) having concave mirrors with high reflectivity in the blue spectral region. The free spectral range (FSR) of the FPI is 1 034 MHz. The spatial distribution of the blue light transmitted through the interferometer is monitored with a CCD camera to optimize the coupling into the fundamental $TEM_{00}$ mode using a combination of lenses.

Finally, an acousto-optic modulator (AOM) is employed to modulate the CBL intensity entering the FPI so that its ultimate spectral resolution can be determined from the transient transmitted response.

## 3. Results

CBL is generated by a resonant FWM process in atomic media with a diamond-shaped transition configuration (Fig. 1a); however, only two applied laser fields are necessary. In the case of Rb atoms the lasers drive the $5S_{1/2} \rightarrow 5P_{3/2}$ and $5P_{3/2} \rightarrow 5D_{5/2}$ transitions to produce a population inversion on the $5D_{5/2} \rightarrow 6P_{3/2}$ transition, generating the third field at 5.23 μm. Mixing of the applied laser fields and this mid-IR field produces optical radiation at 420 nm in the direction satisfying the phase-matching relation $\boldsymbol{k}_{BL} = \boldsymbol{k}_1 + \boldsymbol{k}_2 - \boldsymbol{k}_{IR}$, where $\boldsymbol{k}_{BL}$, $\boldsymbol{k}_1$, $\boldsymbol{k}_2$ and $\boldsymbol{k}_{IR}$ are the wave vectors of the radiation at 420, 780, 776 nm and 5.23 μm, respectively. In contrast to conventional optical parametric oscillation it is not necessary to use an optical resonator to enhance the process: the Rb vapour provides not only the nonlinear medium, but also sets resonant frequencies for new fields through the phase matching relation.

### 3.1. CBL single-mode regime

An overall picture of the CBL spectra is obtained by scanning the 780 nm laser frequency through the $^{85}$Rb $5S_{1/2}(F=3) \rightarrow 5P_{3/2}$ transitions while the 776 nm laser is held at the fixed frequency that produces the highest maximum CBL intensity as the 780 nm laser is scanned.

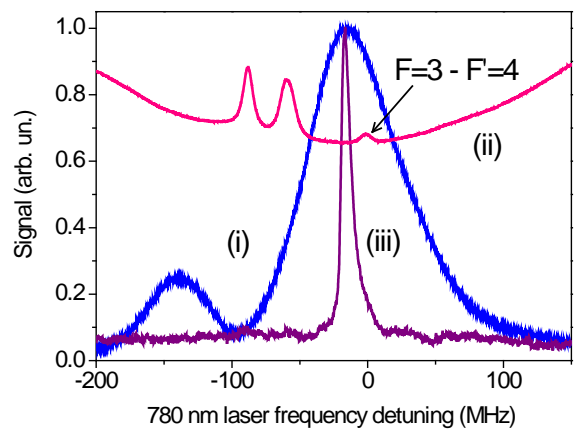

Figure 2. (i) Normalized CBL power, (ii) saturation-spectroscopy signal on the $^{85}$Rb $5S_{1/2}(F=3) \rightarrow 5P_{3/2}$ transitions and (iii) the blue FPI transmission resonance as a function of frequency detuning of the 780 nm laser from the cycling F=3→F′=4 transition. The unlabelled saturation-spectroscopy peaks are Doppler-free crossover absorption resonances. The 776 nm laser frequency is tuned to maximise the CBL.

At an intermediate atomic density ($N \approx 8\times10^{11}$ cm$^{-3}$) this CBL maximum occurs when the 780 nm is tuned close to the 5S$_{1/2}$(F=3)→5P$_{3/2}$(F'=4) transition, as shown in Fig. 2. A precise frequency reference is provided by a Doppler-free saturated-absorption spectrum taken simultaneously in the auxiliary cell. In this situation the intensity-dependent CBL profile is highly asymmetric [4]. The length of the FPI was adjusted so that the transmission peak of the fundamental TEM$_{00}$ mode coincided with the maximum of the CBL profile. Results of our investigation of the absolute frequency of the CBL are presented in [12], where we found that the frequency of the CBL is tunable over a range of approximately 250 MHz, considerably smaller than the Doppler width of the $^{85}$Rb D$_2$ transition, and centred at the $^{85}$Rb 5S$_{1/2}$(F=3)→6P$_{3/2}$(F'=4) transition frequency.

The spectral width of CBL radiation was measured using the blue FPI in scanning mode and with the frequencies of both lasers locked by the method described above. Figure 3a shows the CBL transmission through the scanned FPI while the laser frequencies are locked. The spectral interval between two highest peaks, which are due to coupling of the CBL to TEM$_{00}$ interferometer modes, corresponds to the FSR (1 034 MHz) of the interferometer. The next largest transmission peak represents coupling to the TEM$_{20}$ and TEM$_{02}$ modes, identified by spatial intensity distributions using the CCD camera. Smaller transmission peaks are due to coupling to higher-order transverse modes of the blue FPI and provide a convenient finer frequency scale.

Figure 3b shows a single TEM$_{00}$ transmission resonance. The smooth curve is a Lorentzian fit with a 1.3 MHz width (FWHM). This value represents an upper limit to the CBL linewidth, as it includes technical contributions from the instrumental width of the FPI and the laser linewidths, and a more fundamental contribution from the wave mixing process itself. We next consider each of these contributions in turn.

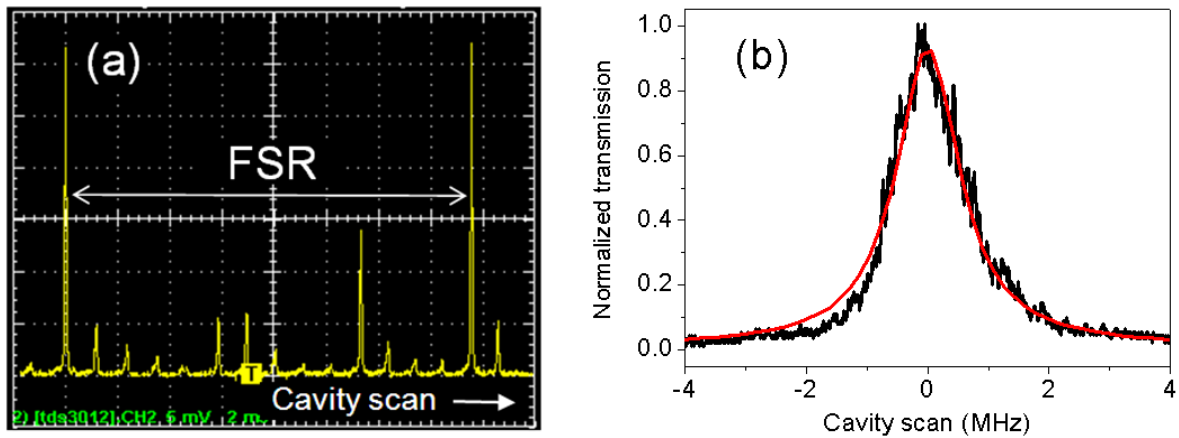

Figure 3. FPI transmission resonances taken when the interferometer scan is (a) larger than the FSR and (b) 8 MHz wide (black curve: transmission profile for TEM$_{00}$ mode; red curve: Lorentzian fit with 1.3 MHz FWHM). The 780 nm laser is locked to the $^{85}$Rb 5S$_{1/2}$(F=3)→5P$_{3/2}$(F'=4) transition while the 776 nm laser is tuned to maximum CBL and locked to the reference cavity.

*3.2.   Interferometer contribution to the CBL linewidth*
We have estimated the intrinsic bandwidth of the blue FPI by measuring the temporal response to input light pulses. The transient response of an optical interferometer with small losses is exponential with a time constant that depends on the round-trip time and losses, but not significantly on acoustic vibrations. More importantly, the time constant is independent of the spectral purity of the input radiation [15].

To rapidly vary the CBL input intensity, the intensity of the 776 nm drive laser is modulated by driving an AOM with square-shaped pulses. The obtained decay time $\tau_0$ of approximately 0.56 $\mu s$ implies a bandwidth $\Delta\nu_{FP} = 1/(2\pi \tau_0)$ and contribution to the width of the observed CBL transmission resonances from the FPI of only 280 kHz. We note that the observed cavity decay time is not sensitive to acoustic vibrations and drift of the interferometer mirrors, while the observed CBL linewidth is. Hence the measured bandwidth should be considered as a lower limit for the FPI contribution to the observed CBL linewidth.

*3.3. Atomic density dependence of CBL linewidth*

Although we expect that the spectral properties of CBL depend on the conditions within the Rb cell, we find that significant atomic density variations, at least in the regime of the present experiment, do not noticeably affect the width of FPI resonances and thus the CBL linewidth. We also studied the CBL linewidths for the two isotopes in the natural mixture cell, for which the atomic densities differ by a factor of $N_{85}/N_{87} \approx 2.7$. Figure 4 shows that the FPI transmission resonances obtained at the same cell temperature on the strongest cycling transitions of the $^{85}$Rb and $^{87}$Rb atoms reveal similar spectral widths despite the different CBL intensities. This intensity difference (a factor of ~20) is much higher than the atomic density difference, probably due to the threshold effect for the generation of mid-IR radiation.

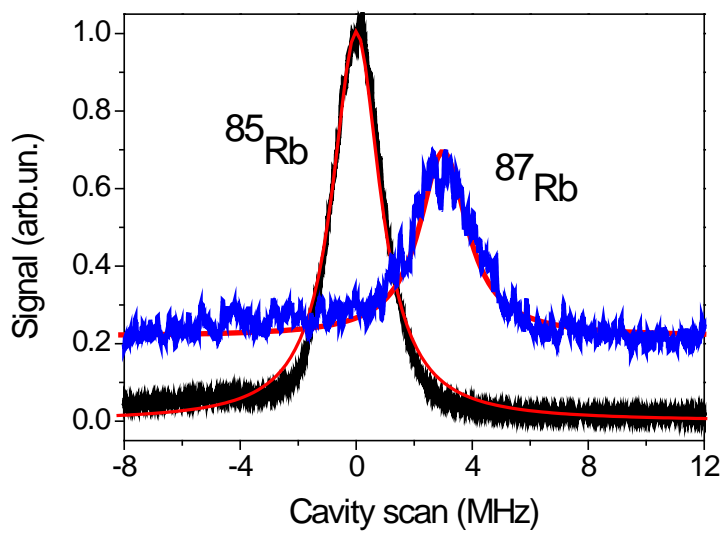

Figure 4. FPI transmission resonances of the CBL from the two Rb isotopes observed under the same experimental conditions except for the frequency of the 780 nm laser, which is locked to the $^{85}$Rb $5S_{1/2}(F=3)\to5P_{3/2}(F'=4)$ or to the $^{87}$Rb $5S_{1/2}(F=2)\to5P_{3/2}(F'=3)$ transition, while the frequency of the 776 nm laser is unchanged. The FPI length is adjusted for maximum transmission in each case, though the scanning range is the same. The $^{87}$Rb signal has been multiplied by a factor of 10, and shifted vertically for clarity. The red smooth curves represent fits of Lorentzian functions of the same width.

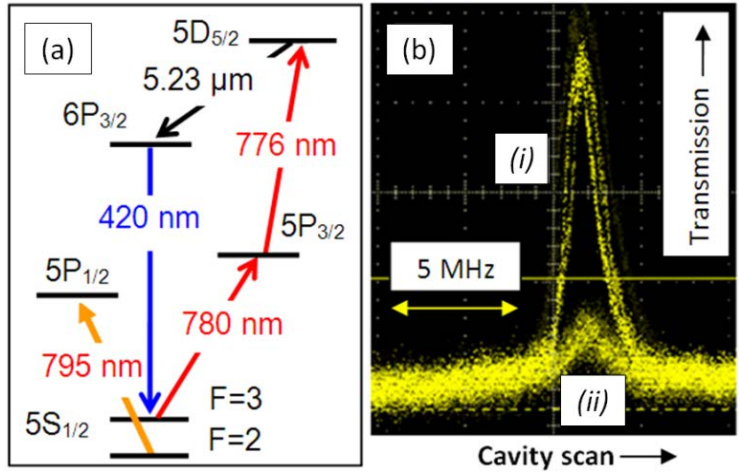

Figure 5. (a) Scheme of hyperfine optical pumping at 795 nm in $^{85}$Rb.
(b) FPI transmission resonance (*i*) with and (*ii*) without optical pumping radiation at 795 nm applied to the $^{85}$Rb $5S_{1/2}(F=2)\rightarrow 5P_{1/2}(F'=3)$ transition. The 780 nm laser is locked to the $5S_{1/2}(F=3)\rightarrow 5P_{3/2}(F'=4)$ transition, while the 776-nm laser is tuned for maximum CBL power.

The number of atoms in the resonant velocity group involved in CBL generation can also be enhanced by optical pumping. Hyperfine optical pumping with an additional laser may cause a dramatic increase in the CBL intensity, particularly when it occurs just above threshold conditions for the CBL generation [14].

We have compared the position and shape of the FPI resonance with and without optical pumping. To minimize effects associated with the interferometer temperature and pressure drifts, the optical pumping laser beam was chopped mechanically at ~200 Hz, synchronised with the interferometer sweep frequency. A digital oscilloscope recorded the FPI resonances with and without optical pumping with a time separation that makes the typical interferometer thermal and pressure drifts negligibly small. We have found that the CBL linewidth remains unchanged within the resolution of our experiment despite an approximately five-fold CBL-intensity enhancement, as shown in Fig. 5b. These observations also highlight the ability to control the CBL intensity through optical pumping.

Similar experiments with fast intensity modulation of the applied laser fields at 780 and 776 nm performed using an AOM indicate that within our experimental resolution the CBL linewidth does not change when intensity variations of more than 50% occur in the applied laser fields. We find that there are no measurable variations in the CBL linewidth over the entire blue-light tuning range.

### 3.4. *Laser linewidth contribution*
To evaluate the contribution to the CBL linewidth from the applied laser light we have estimated the linewidth of the 780 nm and 776 nm lasers by using an optical heterodyning method. The light from a narrow-linewidth reference laser was mixed consecutively with the light from the lasers used for the CBL generation. The beat signals detected by a fast photodiode were recorded on an RF spectrum analyser. We find that for a 100 ms integration time interval the linewidths (FWHM) are approximately 1.0 and 0.6 MHz, respectively. Thus, the estimated laser linewidth contribution of approximately 1.2 MHz is close to the measured width of the typical blue FPI transmission resonances.

A link between the spectral width of FWM-generated signals and laser linewidth was previously demonstrated. The spectral broadening of FWM-generated emission that is additional to broadening from the lasers in semiconductor devices was reported in [16,17] and found to be determined by spectral properties of the laser frequency noise. This broadening is at the sub-kHz level in the case of nearly-degenerate atomic media excited by mutually very coherent optical fields prepared from same laser [18,19].

In the case of atomic media with diamond-shaped transition configurations, studying such broadening at the sub-MHz level requires highly coherent lasers. However, instead of narrowing the laser spectral line further, which is a technically challenging task at such linewidth levels [13], a correlation between the spectral width of FWM-generated signals and laser linewidth can be shown using variably broadened laser radiation. We have analysed the effect on the blue FPI transmission profiles of frequency modulating the second-step laser at 776 nm tuned to the $5P_{3/2} \rightarrow 5D_{5/2}$ transition. It was found that for high-index 50 kHz frequency modulation the effective width of the CBL follows the laser linewidth, while for high-index, high-frequency modulation the laser sideband splitting leads to splitting in the CBL spectrum as shown in Fig. 6.

In addition, by analysing the CBL frequency noise from intensity fluctuations on the side of the blue FPI transmission peaks, we find that the standard deviation of the CBL frequency is about 200 kHz over a 10 ms interval. This value is close to the sum frequency noise of the applied laser light estimated from the servo system signals. For longer integration intervals the FPI drift contribution becomes dominant.

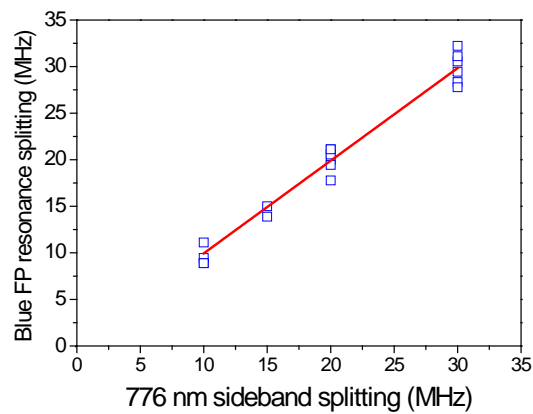

Figure 6. Splitting of the blue FPI transmission profile vs high-index frequency modulation of the 776 nm laser. Red line represents linear fit; slope: 0.995.

Thus we conclude that the CBL spectral width strongly depends on the spectral purity of the applied laser light. Since this limit arises not from atomic relaxation rates or any effect that is intrinsic to the four-wave mixing process itself, but from technical effects, the ultimate linewidth of the generated optical fields may likely be substantially reduced.

*3.5 Linewidth estimation of mid-IR radiation*
Whereas we are presently unable to analyse the spectral content of the mid-IR radiation directly (this will be done in future work) the spectral width of the collimated mid-IR emission at 5.23 µm could be evaluated from the measured width of the typical blue FPI transmission resonances and the measured laser linewidths. Indeed, taking into account that temporal coherences of all optical fields involved into wave mixing process contribute to the new field linewidth, we can conclude that the spectral width of the mid-IR emission is not wider than a few hundred kHz. This means that the linewidth of the internally generated mid-IR radiation is not entirely determined by temporal coherence of the resonant light. Atomic relaxation rates, as well as geometry of the region in which excitation occurs could make even stronger effect on the mid-IR linewidth.

The spectral properties of the internally generated mid-IR radiation at 5.23 µm are important in a number of potential applications, including underwater communication and quantum information processing. The collimation of the backward-directed component of the mid-IR radiation may offer a novel solution [20] to the problem of enhancing the sensitivity of the laser guide star technique [21].

The diamond-shaped transition configuration could also be used for generation of coherent and correlated fields at wavelengths that are difficult to access with other methods.

### 4. Conclusion

We have investigated the linewidth of the collimated blue light at 420 nm generated on the Rb $6P_{3/2} \rightarrow 5S_{1/2}$ transition by a parametric four-wave mixing process in a vapour cell driven by low-power lasers tuned to the $5S_{1/2} \rightarrow 5P_{3/2}$ and $5P_{3/2} \rightarrow 5D_{5/2}$ transitions. The recorded width of transmission resonances of the blue light through a high-finesse Fabry Perot interferometer of 1.3 MHz is accounted for the linewidth of the applied laser fields at 780 nm and 776 nm.

We have found that the CBL linewidth and optical frequency remain unchanged, at least within the resolution of our experiment, despite approximately 270% atomic-density variations and 50% intensity variation of the applied light, and is unchanged over its range of tunability. We expect similar results would be obtained in other alkali atoms.

Because CBL is the product of a wave-mixing process and its linewidth includes contributions from all optical fields involved, then taking into account the measured laser linewidths, it follows that the internally-generated mid-IR light has a linewidth that is no larger than a few hundred kHz and is not determined simply by the applied radiation linewidth. This conclusion can be reached without detecting the 5.23 µm radiation.

Finally, we expect similar high monochromaticity for both wavelength up- and down-converted radiation generated in atomic media with analogous diamond-shape transition configurations if the applied resonant light is highly coherent.

**Acknowledgments**
This work has been supported by the ARC Centre of Excellence for Quantum Atom Optics. AL, CP and GWT thank the ARC for supporting this research through the DP0877938 and FT0990301 research grants. DB is grateful to the Centre for Atom Optics and Ultrafast Spectroscopy at the Swinburne University of Technology for hosting him as a Distinguished Visiting Researcher.